\documentclass[twocolumn]{article}
\usepackage{amsmath, amsthm, amssymb}
\usepackage{graphicx}
\usepackage{hyperref}
\usepackage{url}
\usepackage{rotating,booktabs}
\newtheorem{thm}{Theorem}[section]

\theoremstyle{definition}
\newtheorem{defn}[thm]{Definition}

\theoremstyle{remark}

\numberwithin{equation}{section}

\newcommand{\set}[1]{\left\{#1\right\}}

\begin{document}
\bibliographystyle{plain}

\title{A Survey on Distance Vector Routing Protocols}
\author{Linpeng Tang\footnote{chnttlp@gmail.com}, Qin Liu\footnote{lqgy2001@gmail.com}}

\maketitle
\begin{abstract}
In this paper we give a brief introduction to five different distance vector routing protocols (RIP, AODV, EIGRP, RIP-MTI and Babel) and give some of our thoughts on how to solve the count to infinity problem. Our focus is how distance vector routing protocols, based on limited information, can prevent routing loops and the count to infinity problem.
\end{abstract}
\section{Introduction}
In computer communication theory relating to packet-switched networks, a distance-vector routing protocol is one of the two major classes of routing protocols, the other major class being the link-state protocol. A distance-vector routing protocol uses the Distributed Bellman-Ford algorithm to calculate paths.\cite{WikiDVRouting}

Compared to link-state routing protocols, distance-vector routing protocols only requires the nodes to store next hop and distance information for routing destinations while in a link-state routing protocol, a node has to store all the information about the network topology. And when the metric of a link changes, distance-vector routing algorithms only require the associated node to inform its neighbors while a link state routing protocol requires the node to broadcast the information in the whole network. So typically distance-vector protocols have less computational complexity and message overhead.

However, distance-vector routing protocols also have several disadvantages. The most famous are the routing loop and count to infinity (CTI) problem. Because each router only has limited information about the network topology, routing loops might emerge and lead to CTI problem, greatly impede the efficiency of the protocol. In this survey we will see how several distance vector routing protocols have worked to alleviate or solve this problem. Distance vector routing protocols also suffers from security issues, because routing computation is done distributively, a malfunctioning or malicious node may severely affect the whole network. However, in recent years more secure distance vector routing protocols have been proposed \cite{Hu2003}.Another critical issue is the support for routing areas. In reality, large networks are typically divided into areas to accelerate routing, but distance-vector routing protocols don't support routing areas, so they are not suitable for really big networks.

In general, distance vector routing protocols are more suitable for small or median sized networks or when each node only have a limited storage or computing power. In reality, RIP and EIGRP are two very successful Interior Gateway Routing Protocols and major competitors for the link-state OSPF routing protocol.

Now we'll begin our introduction to RIP, AODV, EIGRP, RIP-MTI and Babel routing protocols. We'll also give some of our thoughts on how to solve the CTI problem(See Section \ref{sec:cti}). In the end we summarize the results in Table \ref{tab:comparison}.

\section{RIP protocol}
The Routing Information Protocol is so far the most popular distance vector routing protocol, and perhaps the most popular interior routing protocol is the TCP/IP suite. RIP protocol's popularity results from its simplicity, early adoption in a popular operating system (BSD) and early standardization in the RFC.

\subsection{Overview of RIP protocol}
RIP protocol is the simplest form of a distance vector routing protocol. For example, the routing table is very simple and only employs very limited information--destination, hop count and next hop.

Its working are also easy to understand. On a regular basis, each router in the network sends out its routing table to its neighbors, informing them to which subnets it is connected and how far these subnets are (by measure of hop count). Once a router receives such a routing message, it updates its routing table. Say router $B$ sends to router $A$ claiming it has a route to $C$ with hop count $K$, then $A$ knows that by going through $B$, it can reach $C$ with hop count $K+1$. RIP protocol is essentially based on a distributed version of the famous Bellman-Ford shortest path algorithm. Assuming the protocol is executed in a synchronous fashion (that is, each round the routers receives the routing message, updates the routing table and sends out a new message, at the same time), then Bellman-Ford algorithm tells us the routing path will converge in no more than $K$ rounds, where $K$ is the diameter of the network.

The RIP protocol has several advantages. It is simple, and it is every efficient for small and simple networks, consuming little network bandwidth and little storage and computing power for the routers (which may be a battery powered small device).

\begin{figure}
  \includegraphics[width=200pt]{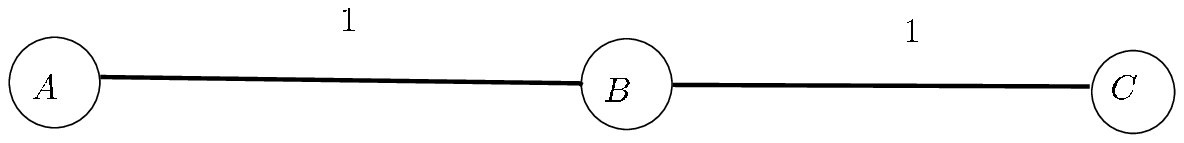}\\
  \caption{a simple network topology that would cause count to infinity problem}\label{pic:RIP-NET1}
\end{figure}

The RIP protocol, however, also suffers from some inherent limitations. The most famous is perhaps the count to infinity problem. In Figure \ref{pic:RIP-NET1} say $A$, $B$ and $C$ are all connected in a network and suddenly $C$ is disconnected because of a corrupted link. We would naturally want $A$ and $B$ to find this out immediately. Now note that $B$ initially has a route to $C$ with hop count 1. Now this route is invalid, but instead $B$ would find that $A$ has a route to $C$ with hop count $2$, and so it assume it has a route to $C$ through A, with a hop count of $3$; then, A would similarly update its route to C with a hop count of $4$; and the process goes on. This kind of routing loop causes the count to infinity problem, draining the network bandwidth, slowing down the routing path convergence and severely impeding the performance of RIP protocol.

RIP partly solves this problem by setting the maximum hop count to be $15$, so a hop count of $16$ means a distance of infinity. And so the above process can at most last for $15$ rounds. Some people say this approach would limit the size of the network that RIP can support. However, a network with a diameter of 15 is actually already too large for RIP protocol (typically routers are organized hierarchically, and just image a tree of routers with a depth of 7 and how many routers would be in this network). The problem is that with such a large network (potentially tens of thousands of computers), the routing table would be quite large, the exchange in routing tables would generate too much traffic and the routing path may converge too slow. In this case, maybe a hierarchical routing protocol is more suitable.

\subsection{RIP Special Features For Resolving RIP Algorithm Problems}
The simplicity of the Routing Information Protocol is its most attractive quality, but in order to solve some of its inherent problems and improve its performance, we have to add to its complexity. However, we will see these can't solve the routing loop and count to infinity completely.\cite{RIPGuide}

\subsection{Split Horizon}
\begin{figure}[htbp]
  \centering
  \includegraphics[width=120pt]{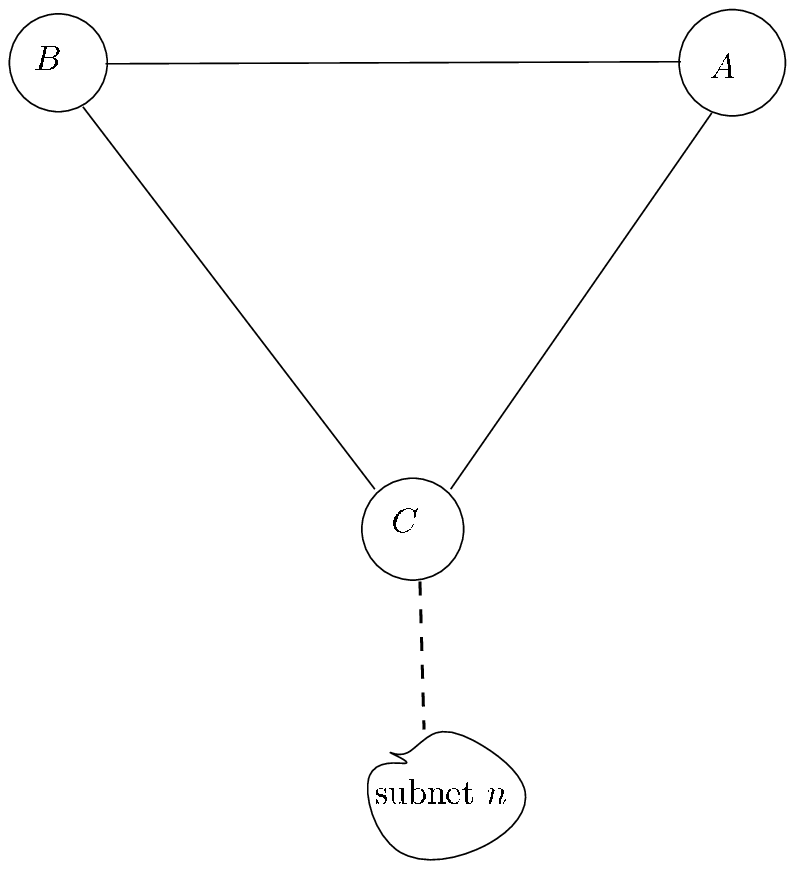}\\
  \caption{A simple network topology where split horizon can't solve the count to infinity problem}\label{pic:RIP-NET2}
\end{figure}

Split Horizon can prevent the routing loop problem in some simple structured networks. And it's intuitively simple. Its rule is that if in $A$'s routing table the next hop to a particular subnet is $B$, then $A$ never reveals to $B$ that it has a route to this subnet. However, this technique can't prevent routing loop in Figure \ref{pic:RIP-NET2}. Now $A$, $B$, $C$ are three routers that are linked to each other and $C$ has a link to a subnet n. And in $A$'s and $B$'s routing table, they both have a route to n with the next hop being $C$. Now if the link from $C$ to $n$ is broken, $A$ would falsely assume that it can reach $n$ from $B$, and so $B$ would similarly assume it can reach $n$ from $A$. So these two nodes would easily form a routing loop.

\subsection{Split Horizon with Poisoned Reverse}
In split Horizon with Poisoned Reverse, if in $A$'s routing table the next hop to a particular subnet is B, then instead of hiding this routing from B, A actively claim its distance to this subnet is infinity (that is $16$ in RIP protocol). This adds more insurance that an outdated routing information would not help form a routing loop in a changing network. However, it can't prevent routing loop as in Figure \ref{pic:RIP-NET2}.

\subsection{Hold-Down}
The hold down feature works by having each router start a timer when they first receive information about a subnet that is unreachable. Until the timer expires, the router will discard any subsequent route messages that indicate the subnet is in fact reachable. A typical hold-down timer runs for $60$ or $120$ seconds. Note that hold-down would actually solve the count to infinity problem in Figure \ref{pic:RIP-NET2} because A and B both refuse to receive updates from each other.

The may disadvantage of hold-down is that it forces a delay in a router responding to a route once it is fixed. in \ref{pic:RIP-NET2}, Suppose a little malfunction causes C to go down for $5$ seconds, but $A$ and B would refuse to route to $n$ for $60$ or $120$ seconds!

\section{AODV routing protocol}
\subsection{Introduction to the AODV protocol}
The Ad-hoc On-Demand Distance Vector(AODV) routing protocol is designed for use in ad-hoc mobile networks. AODV is a reactive protocol, meaning the routes are created only when they are needed\cite{sklyarenko2006aodv}. It assigns each routing reply with a sequence number to determine whether routing information is up-to-date and to prevent routing loops.

Route discovery is based on query and reply cycles. When a node needs a route to a particular node, it sends a routing request to its neighbors and when a route has been found, the reply will be forwarded back to the originator of the request. Each node only needs to remember a set of routing destinations and the next hop of the route with some other auxiliary information, so AODV is a typical distance vector routing algorithm. Four kind of control packets are used: routing request message (RREQ) is broadcasted by a node requiring a route to another node, routing reply message (RREP) is routed back (using the reverse route) to the source of RREQ, and the route error message (RERR) is sent to notify other nodes of the loss of a link, as is common is mobile networks. HELLO messages are sent periodically for detecting and monitoring links to neighbors.

\subsection{Overview of the routing protocol}
The AODV routing protocol is reactive, meaning  that it creates and maintains routes only if they are needed, on demand. Each node will store its own routing table, and every entry in the table includes the following items:
\begin{enumerate}
    \item a routing destination
    \item number of hops of the route
    \item destination sequence number
    \item a list of precursors that may be forwarding packets on the route
    \item the expiration time for this routing table entry
\end{enumerate}
Note that each node performing AODV protocol has a sequence number that is incremented when it sends a RREQ or when a destination node sends back the RREP message. And this sequence number will be used to distinguish the up-to-date route from the older, possibly invalid route , so AODV is completely loop free and copes well in a highly dynamic network. When one node wants to send a message to a node that is not one of its Neighbors, it will broadcast a RREQ message. On receiving the RREQ message, the fellow nodes will try to satisfy the request
\begin{itemize}
\item if its routing table has a valid route to the destination, or it is the destination, it will generate a RREP message back to the originator of the request
\item if it doesn't have a valid route, it will rebroadcast the RREQ message.
\end{itemize}

Note that the RREQ ID will be incremented each time the source node sends a new RREQ, so the pair (source IP address, request ID) identifies a RREQ uniquely. In this way, other nodes can discard the duplicate or the older RREQ messages and saves the communication overhead of RREQ broadcasting.

If a node detects a link breakage to one of its Neighbors(say, a Neighbor has not sent HELLO to it message for a while), it will generate a RERR message. It will first invalidate the existing routes that has the lost Neighbor as the next hop, then lists the affected destinations and determines if any of its neighbors may be affected (such Neighbors may have delivered messages to it using the affected routes), and then delivers an appropriate RERR to such Neighbors.\cite{Perkins2003}

\subsection{Sequence Numbers}
The sequence numbers are the most important feature of AODV for removing the old and invalid information from the network. They works as a sort of time stamps and prevent the AODV protocol from the loop problem\cite{sklyarenko2006aodv}. The destination sequence number for each destination host is stored in the routing table, and is updated in the routing table when the host receives the message with a greater sequence number.

Each node keeps its own sequence number, which is changed in two cases:
 \begin{enumerate}
 \item before it sends RREQ message, the sequence number is incremented.
 \item Immediately before a destination node originates a RREP in response to a RREQ, it MUST update its own sequence number to the maximum of its current sequence number and the destination sequence number in the RREQ packet\cite{Perkins2003} (as sequence number is only 32 bit long, it may wrap to zero after it has reached the maximum value possible, and it is necessary for the destination to always keep its sequence number the largest among the other nodes' opinions, so this step is crucial.)
 \end{enumerate}

\subsection{Solution to the Count to Infinity Problem}
The following proof is excerpted from\cite{sklyarenko2006aodv}.

\begin{thm}
AODV routing protocol is loop-free.
\end{thm}

\begin{proof}
Proof by contradiction: Let $\set{N_1, \dots, N_m}$ be a loop in a route from any source node to any destination node. That means that these nodes are chained to each other. Assume without loss of generality that
\begin{align*}
& nextHop(N_i) = N_{i+1}\\
& \forall 1 \le i \le m, \quad where \quad N_{m+1} = N_1
\end{align*}
Then, from the definition of  AODV destination sequence numbers, we have (Note that node $A$ will only have node $B$ as its next hop if $B$  has a route of same or larger destination sequence number than that of $A$)
\begin{align*}
& DestSeqNo(N_i) \le DestSeqNo(N_{i+1}) \quad \forall i \\
& \Rightarrow DestSeqNo(N_i) = DestSeqNo(N_j) \quad \forall i,j
\end{align*}

This means that the information about the destination node was obtained from the same RREP message. Taking into account of the definition of the hop count, we get
$$hopCount(N_i) = hopCount(N_{i+1})+1$$
But $N_m$ has $N_1$ as its next hop, so
\begin{align*}
hopCount(N_m) & = hopCount(N_1)+m-1\\
 & = hopCount(N_m)+1+m-1
\end{align*}
and we get $m=0$, thus completing the proof.

\end{proof}

\subsection{Evaluation of AODV}
\paragraph{Advantages} AODV has no central administrative system to control the routing process, and reacts very fast to topological changes in the networks. It saves storage and energy since each node only needs to store the routing entries which it is responsible or interested in.

\paragraph{Disadvantages} It's possible that a valid route is expired. And it is often hard to determine a reasonable expiration time--a too short ET will cause repeated, unnecessary routing requests while a too long ET will cause the protocol slow to changes. And since AODV only stores a very limited information in each node(destination, next hop), each node has a limited communication range, this causes AODV to rely on route discovery flood more often, which may carry significant network overhead. The performance of AODV protocol is poor in larger networks, since larger networks has longer paths and longer paths is more vulnerable to link breakages and requires high control overhead (may cause many more broadcasts) for its maintenance.\cite{sklyarenko2006aodv} Moreover, AODV is vulnerable to various kinds of attacks, because it is based on the assumption that all nodes will cooperate.

\section{Enhanced Interior Gateway Routing Protocol}
Enhanced Interior Gateway Routing Protocol(EIGRP) \label{sec:eigrp} is a Cisco proprietary routing protocol loosely based on their original IGRP. EIGRP is an advanced distance-vector routing protocol, with optimizations to minimize both the routing instability incurred after topology changes, as well as the use of bandwidth and processing power in the router. The basis of its operation is the Diffusing Update Algorithm (DUAL)\cite{garcia1993loop}, which is used to compute shortest paths distributedly and without ever creating routing-table loops or incurring counting-to-infinity behavior.

\subsection{Overview of EIGRP}
In this section, we will introduce how EIGRP operates.

\subsubsection{Diffusing Update Algorithm}
This section provides a summary of the operation of DUAL as it applies in EIGRP, the details can be found in \cite{albrightson1994eigrp}.

Each router maintains a vector with its distance to every known destination in the routing table.
Routing information is exchanged only between neighbors by means of update messages; this is done after
routers detect changes in the cost or status of links. Each update message contains a distance vector of
one or more entries, and each entry specifies the length of the selected path to a given destination, as well
as an indication of whether the entry constitutes an update, a query, or a reply to a previous query. A
router also maintains a topology table containing the distance reported by
selected neighbor routers to each known destination. Information for the routing table is taken from the
topology table.

For a given destination, a router updates its routing table differently depending on whether it is
\emph{passive} or \emph{active} for that destination. A router that is passive for a given destination can update the
routing-table entry for that destination independently of any other routers, and simply chooses as its new
distance to the destination to be the shortest distance to that destination among all neighbors, and as
its new successor to that destination to be any neighbor through whom the shortest distance is achieved.
In contrast, a router that is or becomes active for a given destination must synchronize the updating
of its routing-table entry with other routers. A router is active if it is waiting for at least one neighbor
to send a reply to a query already sent by the router, and is passive otherwise. Furthermore, a router
is initialized in passive state for all known destinations with a $0$ distance to itself and a finite distance
to other destinations that are directly attached to an adjacent link. Passive destinations with infinite
distances are removed from the topology table.

When a router is passive and needs to update its routing table for a given destination $j$ after it
processes an update message from a neighbor or detects a change in the cost or availability of a link, it
tries to obtain a successor. From router $i$'s standpoint, a successor toward destination $j$
is a neighbor router $k$ that satisfies the following two equations:
\[D_j^i = D_{j k}^i + c_k^i = \min \{D_{j p}^i + c_p^i | \text{p is a neighbor}\}\]
\begin{equation}
\label{equ:fc}
D_{j k}^i < FD_j^i
\end{equation}
where:\\
$D_j^i = $ current distance from router $i$ to destination $j$\\
$D_{j k}^i = $ distance to destination $j$ reported by neighbor $k$ as known by router $i$\\
$c_k^i = $ cost of the link to neighbor $k$ as known by router $i$\\
$FD_j^i = $ feasible distance for destination $j$, and equal to the minimum value obtained for $D_j^i$ since the last
time router $i$ transitioned from active to passive state for destination $j$.

If router $i$ finds a successor, it remains passive and updates its routing table entry.
Alternatively, if router $i$ cannot find a successor, it first sets its distance equal to the addition
of the distance reported by its current successor plus the cost of the link to that neighbor. The router also
sets its feasible distance equal to its new distance. After performing these updates, the router becomes
active by sending a query in an update message to all its neighbors; such a query specifies the router's
new distance through its current successor.

Once active, a router cannot change its successor, its feasible distance, the value of the distance it
reports to its neighbors, or its entry in the routing table, until it receives all the replies to its query. A
reply received from a neighbor indicates that such a neighbor has processed the query and has either
obtained a successor to the destination, or determined that it cannot reach the destination. Once
Node $i$ obtains all the replies to its query, it computes a new distance and successor to destination $j$,
updates its feasible distance to equal its new distance, and sends an update to all its neighbors.

\subsubsection{Feasibility Condition}
The feasibility condition given by Equation (\ref{equ:fc}) is called \emph{source node condition}(SNC). Using SNC when nodes choose their successors is sufficient to select minimum cost loop-free paths.

To show that the SNC guarantees loop-freeness, recall that at the time when router $i$ accepts an update from router $k$, $D_{j k}^i$ announced by router $k$ is no smaller than $FD_j^k$; since it is smaller than $FD_j^i$, at that point in time $FD_j^k < FD_j^i$. Since this property is preserved when router $i$ sends updates, it remains true at all times, which ensures that there are no loops.

The full proof of SNC can be found in \cite{garcia1993loop}.

\subsubsection{Feasible Successor}
EIGRP also uses \emph{Feasible Successor} to improve its performance. A feasible successor for a particular destination is a next hop router that satisfies Feasibility Condition.

A feasible successor provides a working route to the same destination, although with a higher distance than a successor. At any time, a router can send a packet to a destination marked "Passive" through any of its successors or feasible successors without alerting them in the first place, and this packet will be delivered properly. Feasible successors are also recorded in the topology table.

The feasible successor effectively provides a backup route in the case that existing successors die. Also, when performing unequal-cost load-balancing (balancing the network traffic in inverse proportion to the cost of the routes), the feasible successors are used as next hops in the routing table for the load-balanced destination.

By default, the total count of successors and feasible successors for a destination stored in the routing table is limited to four.

\subsubsection{Composite and Vector metrics}
EIGRP, like IGRP represents distances as a composite of available bandwidth, delay, load utilization, and link reliability.

For the purposes of comparing routes, the vector metrics are combined together in a weighted formula to produce a single overall metric\cite{wiki:eigrp}:
\begin{align*}
& \bigg [ \bigg ( K_1 \cdot \text{Bandwidth} + \frac{K_2 \cdot
\text{Bandwidth}}{256-\text{Load}} + K_3 \cdot \text{Delay} \bigg )\\
& \cdot \frac {K_5}{K_4 + \text{Reliability}} \bigg ] \cdot 256
\end{align*}
where the various constants (K1 through K5) can be set by the user to produce varying behaviors. An important and totally non-obvious fact is that if K5 is set to zero, the term is not used (i.e. taken as 1).

The scaling factor, 256, was introduced as a simple means to facilitate backward compatility between EIGRP and IGRP.

EIGRP also maintains a hop count for every route, however, the hop count is not used in metric calculation. It is only verified against a predefined maximum on an EIGRP router (by default it is set to 100). Routes having a hop count higher than the maximum will be advertised as unreachable by an EIGRP router.

\subsection{Evaluation}
\subsubsection{Advantages}
\begin{enumerate}
\item Accurately routing load calculating. EIGRP uses composite metrics.
\item Loop-free and fast convergence. EIGRP uses DUAL , only routing table changes are propagated; and to one route ,only relative routers will recalculates.
\item Low usage of network resource. During normal operation, usage of network resource is very low; only hello packets are transmitted on a stable network. When a change occurs, only routing table changes are propagated, not entire routing table; this reduces the load the routing protocol itself places on the network.
\end{enumerate}

\subsubsection{Disadvantages}
\begin{enumerate}
\item There is no area in EIGRP, so it is not good at dealing with big hierarchy network.
\item In some cases, the routers will be active for quit a long time, this affect the fast convergence seriously. For instance, in a long and narrow network, if something has changed, it would take EIGRP a long time to send the message from one side to the other side.
\item In a broadcast network, EIGRP sets up a full mesh adjacency relationship with each other, the routers exchange information with other. This would waste a lot of bandwidth.
\item EIGRP is a protocol come up with by Cisco, it is a private protocol, not a open standard.
\end{enumerate}

\section{RIP-MTI protocol}

\subsection{introduction to the RIP-MTI protocol}
RIP is popular for its simplicity, but it also suffers from the CTI problems most severely. Is there any way to design a better routing protocol that is compatible with the widely used RIP protocol and solves the CTI problem simultaneously? This sounds like a impossible mission, however, according to \cite{Steigner2008}, by exploiting the distance vector updates more thoroughly than common RIP protocols, it is possible to design a much better protocol, and this new protocol is called Routing Information Protocol with Minimal Topology Information (RIP-MTI). In this section we will see how the new protocol has exploit the routing minimal information communicated in the RIP protocol and discuss its advantages and limitations.

\subsection{The Network Model}
First we introduce some basic terms like hop, path, Simple Loop and Source Loops that are used by the designers of the RIP-MTI protocol. The follow definitions are from \cite{Steigner2008}.

A network consists of subnets and routers which are connected via interfaces. Therefore we have a set of subnets $SN = \set{s_1,\dots,s_n}$, and a set of interfaces $IF = \set{if_{1}, \dots, if_{m}}$ which are used to define the set of routers $R = \set{r_1,\dots, r_k}$. A router $r\in R$ is identified by its interfaces $r=\set{A, B, C, \dots}$, $r\subset IF$ and $\forall A \in IF, (A\in r_i\Rightarrow A\not \in r_j)$ for $r_i, r_j \in R, i\neq j$. That is, an interface is an element of one unique router only. Interfaces are connected to subnets and the topology of a network is given by the relation $CON\subset IF\times SN$ where $(if_i, s_j) \in CON$ if and only if interface $if_i$ is connected to subnet $s_j$.

An elementary step from a router $i\in R$ via outgoing interface $O\in i$ to an adjacent router j via incoming interface $I\in j$ using subnet $s\in SN$ is called a hop and is defined by a 3-tuple $H_{i,j} = (O, s, I)$, where $(O, s), (I, s) \in CON$. If the destination of a hop is simply given by a subnet and not a designated interface of a router, we write it as $H_{i,s} = (O,s, *)$.

A path through the network is a sequence of hops and the metric of a path is the number of hops this sequence consists of. For example, the path $P_A^{i,d}$ beginning at router $i \in R$ leading to a subnet $d \in SN$ is given by
\begin{align*}
P_A^{i,d}
&= (H_1, H_2, \dots, H_l) \\
&=((O_1, s_1, I_1), (O_2, s_2, I_2), \dots (O_l, s_l, *))
\end{align*}
\begin{figure}[htbp]
  \includegraphics[scale=.4]{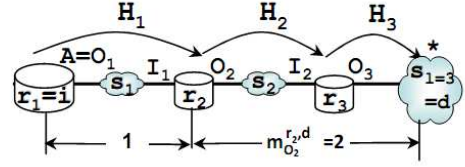}\\
  \caption{A particular from router $i$ to subnet $d$ with a metric of $3$}\label{pic:mti-network}
\end{figure}

Now we introduce the defintion of Simple Loop and Source Loop, which are of special interest to RIP-MTI protocol.

\begin{defn}
A Simple Loop is a path $P_{A,B}^{i,d,i}$ where $A,B \in i, O_1 =A, I_l=B, \exists n 1\leq n \leq l, s_n = d$, and $\forall I_j 1\leq j < l, I_j \not \in i$.
\end{defn}
Intuitively, simple path goes from one interface to a router to the other interface of the same router without going through it in between.

\begin{defn}
A Source Loop is a path $P_{A,B}^{i,d,i}$ where $A, B\in i, O_1=A, I_l=B, \exists n 1\leq n \leq l (s_n=d)$, and $\exists I_j 1\leq j < l$, $I_j\in i$.
\end{defn}
So, unlike a Simple Loop, a Source Loop starts from a router, goes through it in the middle and ends in the same router at a different interface, note that we don't care which interfaces the path intersects with the router in the middle, and even the starting interface or the finishing interface will do.

Furthermore, Source Loops are classified into two basic types. Lets denote the Source Loop as $(A\in R_1, b_1, O_1),\dots, (I_{j-1}, b_{j-1}, U\in R_1), (V\in R_1, b_j, O_{j}), \dots, (I_m\in R_1, b_m, V\in R_1)$. So $A, U, V, B$ belong to the same router.
\begin{itemize}
  \item X-Combination: $V \neq B$
  \item Y-Combination: $V = B$
\end{itemize}
We will use different methods to detect X-Combination Source Loops and Y-Combination Source Loops later.

\subsection{Loop Detection}
Loop detection lies at the heart of RIP-MTI protocol. In a complex network, there may be many loops. Now if there is a simple loop $P_{A,B}^{i,d,i}$ that starts from router $i$ at interface $A$, goes to subnet $d$ and comes back at interface $B$, then it means there are at least two routes from $i$ to $d$, and if one route starting from $A$ is corrupted, we may switch to the route starting from $B$. However, if $P_{A,B}^{i,d,i}$ is a Source Loop, then a harmful routing loop might form if we accept the routing updates from $B$. So the critical question is to distinguish between Simple Loops and Source Loops.

The minimum Simple Loop metric (msilm) between to interfaces $A$ and $B$ on router $i$ is:
$$msilm_{A,B}^i = \min\set{silm_{A,B}^{i,d,i} for all subnets d}$$
where $silm_{A,B}^{i,d,i}$ means the metric of a Simple Loop $P_{A,B}^{i,d,i}$.

Further on we define the minimal return path metric (mrpm) as follows:
\begin{align*}
& mrpm_A^i = \min \{silm_{A,B}^{i,d,i} \quad\\
& \text{for all interfaces $B\neq A$ of router $i$}\}
\end{align*}

Now we have the following two sufficient conditions for detecting Source Loops:
\begin{itemize}
  \item A path $P_{A,B}^{i,d,i}$ with metric $m_{A,B}^{i,d,i}$ is no X-Combination, if the following inequality holds:
    $$m_A^{i,d}+m_B^{i,d}-1 < mrpm_A^i + mrpm_B^i$$

  \item A path $P_{A,B}^{i,d,i}$ with metric $m_{A,B}^{i,d,i}$ is no Y-Combination, if the following inequality holds:
    $$m_A^{i,d} < mrpm_A^i+m_B^{i,d}$$
\end{itemize}
Please note that \emph{even a path fails for these paths, it may still be a valid path} (because we only have such limited information at hand, we can't have a necessary-and-sufficient condition for source loop detection).

\subsection{RIP-MTI routing}
\subsubsection{Loop detection}

Suppose that router $i$ has an entry to subnet $d$ via interface $B$ with metric $m_B^{i,d}$. This indicates a path $P_B^{i,d}$. Now $i$ receives an update to subnet $d$ via interface $A\neq B$. This indicates an alternate path $P_A^{i,d}$ to subnet $d$ with metric $m_A^{i,d}$. The combination of these path results in a path $P_{A,B}^{i,d,i}$ with metric $m_{A,B}^{i,d,i}=m_A^{i,d}+m_B^{i,d}-1$. We want to know if $P_{A,B}^{i,d,i}$ is a Simple Loop or a Source Loop. And to achieve this, we check the metrics on X- and Y-Combination by the two inequalities we have established. If the path passes the test, it is a Simple Loop, and we will upate the minimum Simple Loop metric and the minimum return path metric accordingly.

\subsubsection{CTI Situations}
Now we briefly introduce how RIP-MTI avoids routing loops. For simplicity we assume the network routing path has converged. Now consider a typical situation when a routing loop might occur. Router $i$ has a path to subnet $d$ from interface $B$, but now the link is corrupted, and $i$ receives an update from $B$ which makes the entry invalid. And then $i$ receives an update from $A$ claiming an alternative path to $A$. How can $i$ decide whether this update is valid or not? It uses the loop information in the mslm and mrpm table.
\begin{itemize}
\item If there is no simple path between $A$ and $B$, the update must be rejected. (No simple path means no alternative path)
\item If $m_A^{i,d} + m_B^{i,d} - 1 < mslm_{A,B}$, the update will be rejected.
\item $m_A^{i,d} \geq mrpm_A^{i,d} + m_B^{i,d}-1$, the update will be accepted or rejected according to the working modes, because we are not sure whether it's a valid update.
\end{itemize}

Please note that there are still many details and subtleties that I omit in this gentle introduction, and the interested reader may consult \cite{bohdanowicz2009avoidance}, which provides a detailed explanation. There are actually three working modes in RIP-MTI:
\begin{enumerate}
\item The Normal Mode rejects only alternative routes which
have been verified as invalid.
\item The Strict Mode accepts only the alternative routes
which have been verified as valid.
\item The Careful Mode accepts only the alternative routes
which have been verified as valid and marks the invalid
for further consideration.
\end{enumerate}

\subsection{Summary of RIP-MTI protocol}
According to \cite{Steigner2008}, the designers of the protocol has tested the RIP-MTI protocol using the visualization techniques, and the results show that RIP-MTI does converge faster than RIP in complex network topologies.

In summary, RIP-MTI is indeed a very interesting distance vector routing protocol. It is downward compatible with the simple RIPv2 protocol and at the same time overcomes the shortcomings of RIP using only very limited information. The key point is that it traces the different routing updates coming from different interfaces of one router, and using this information it can learn something about the network topology and thus avoid many CTI problems.

The downside of RIP-MTI, of course is that it still cannot eliminate CTI problems completely and soundly--by completely I mean rejecting any invalid routing updates, and by soundly I mean accepting any valid updates. The normal mode will accept some invalid updates and the strict mode will reject some valid updates. The careful mode somehow provides a compromise between the two.

The RMTI project has implemented RIP-MTI protocol included a whole ``Qugga Routing Suite''. And we will see whether this new approach will gain its popularity.

\section{Babel routing protocol}
The Babel routing protocol is a mostly loop-free distance-vector routing protocol that is designed to be robust and efficient on both wireless mesh networks and classical wired networks. It is based on the idea in DSDV\cite{perkins1994highly} and EIGRP in Section \ref{sec:eigrp}.

\subsection{Overview of Babel}
In this section, we describe how Babel operates conceptually and we only detail the part that are different from DSDV and EIGRP.

\subsubsection{Feasibility Condition}
Babel uses the same feasibility condition, SNC, that is used in EIGRP.

\subsubsection{Solving Starvation: Sequencing Routes}
Obviously, the feasibility conditions defined above cause starvation when a router runs out of feasible routes. Consider the following diagram, where both A and B have selected the direct route to S:
\begin{center}
  \includegraphics[scale = 0.4]{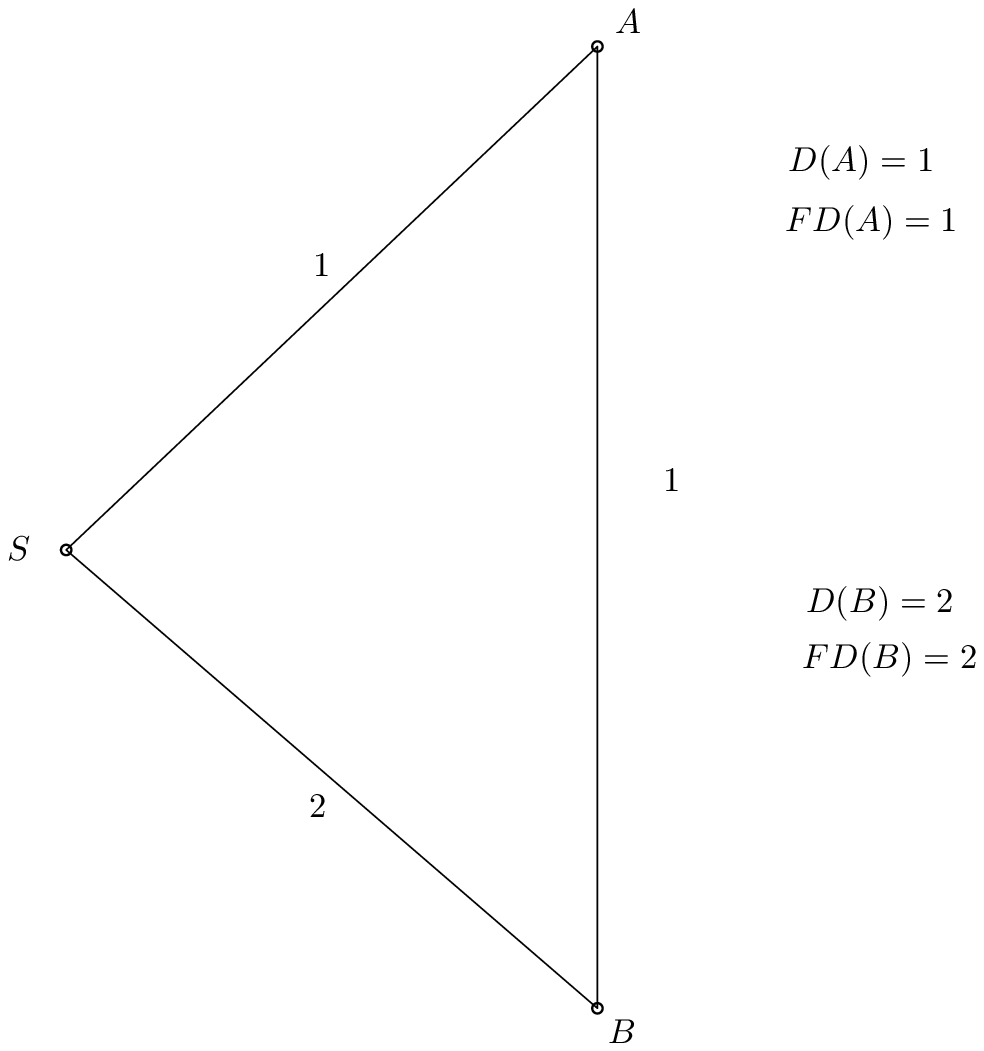}
\end{center}
Suppose now that the link between A and S breaks:
\begin{center}
  \includegraphics[scale = 0.4]{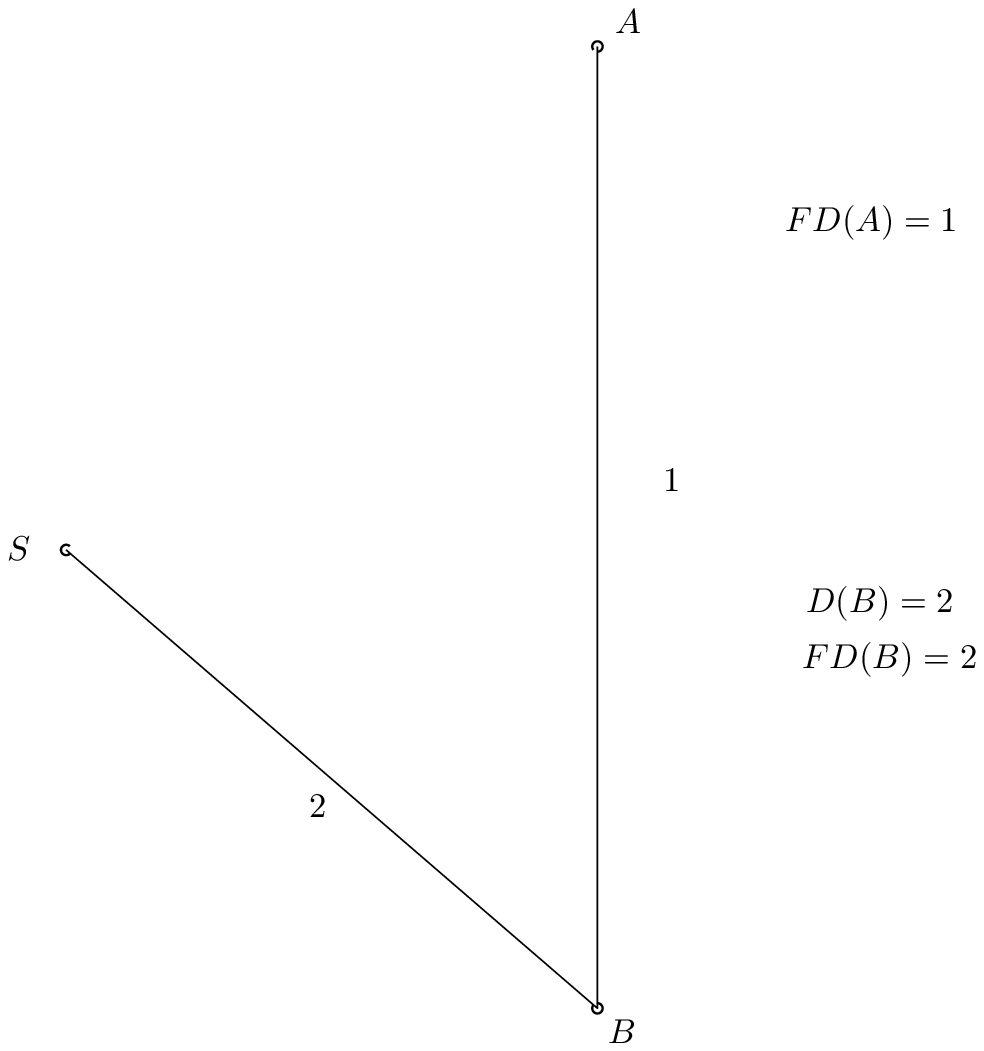}
\end{center}
The only route available from $A$ to $S$, the one that goes through $B$, is not feasible: $A$ suffers from a spurious starvation.

At this point, the whole network must be rebooted in order to solve the starvation; this is essentially what EIGRP does, when it performs a global synchronization of all the routers in the network with the source (the "active" phase of EIGRP).

Babel reacts to starvation in a less drastic manner, by using sequenced routes, a technique introduced by DSDV and adopted by AODV. In addition to a metric, every route carries a sequence number, a nondecreasing integer that is propagated unchanged through the network, and is only ever incremented by the source; a pair $(s, m)$, where s is a sequence number and m a metric, is called a distance.

A received update is feasible when either it is more recent than the feasibility distance maintained by the receiving node, or it is equally recent and the metric is strictly smaller. More formally, if $FD(A) = (s, m)$, then an update carrying the distance $(s', m')$ is feasible when either $s' > s$, or $s = s'$ and $m' < m$.

Assuming the sequence number of $S$ is $137$, the diagram above becomes:
\begin{center}
  \includegraphics[scale = 0.4]{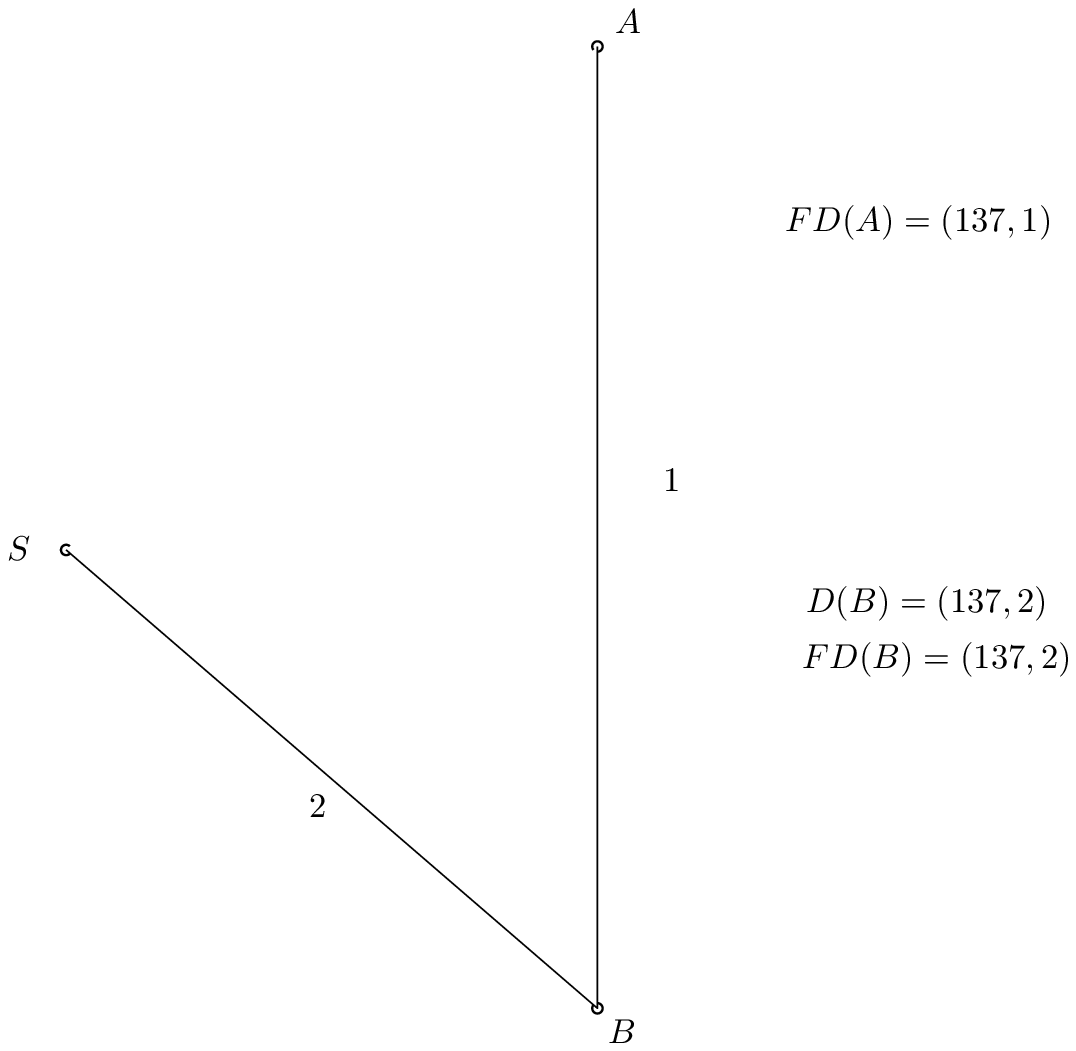}
\end{center}
After S increases its sequence number, and the new sequence number is propagated to B, we have:
\begin{center}
  \includegraphics[scale = 0.4]{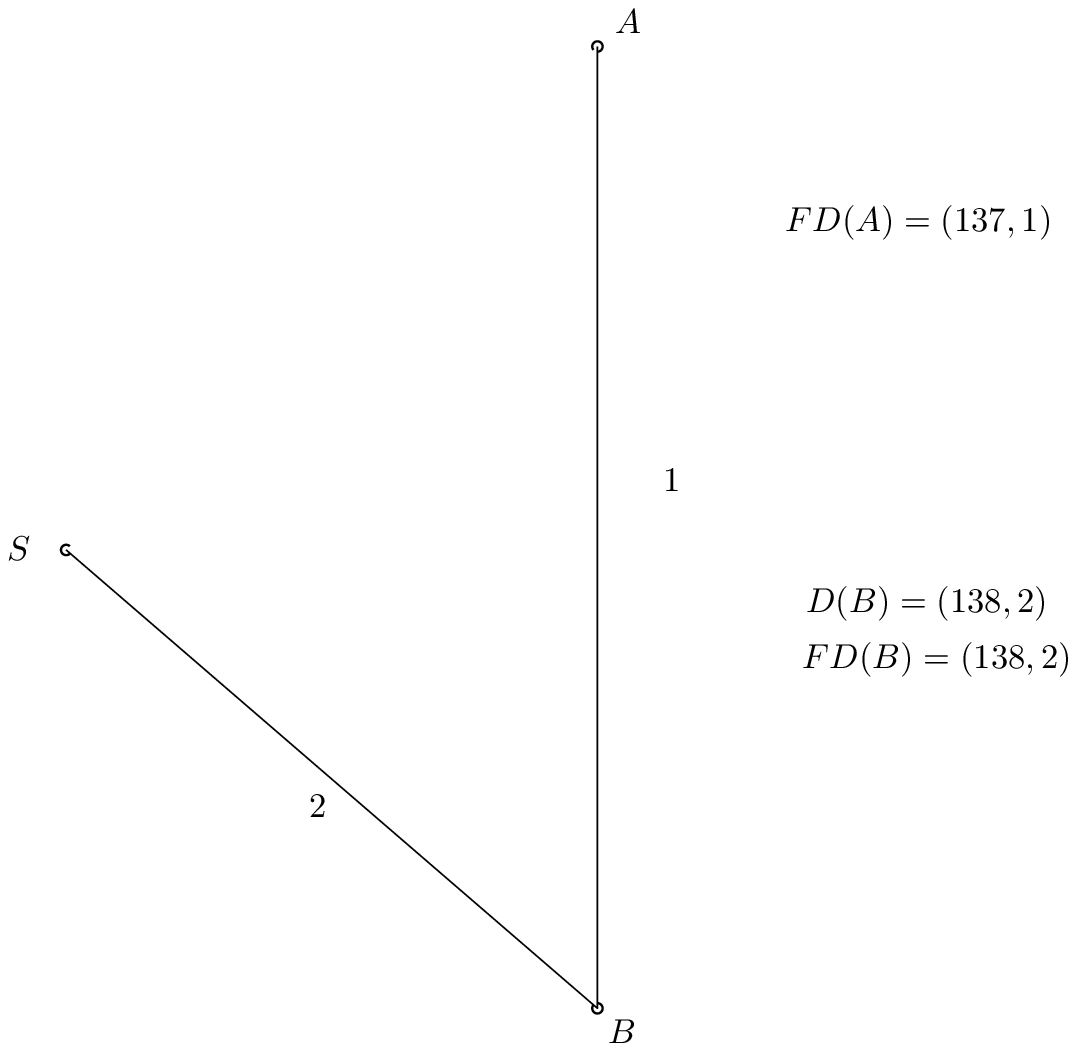}
\end{center}
at which point the route through B becomes feasible again.

\subsubsection{Requests}
In DSDV, the sequence number of a source is increased periodically. A route becomes feasible again after the source increases its sequence number, and the new sequence number is propagated through the network, which may, in general, require a significant amount of time.

Babel takes a different approach. When a node detects that it is suffering from a potentially spurious starvation, it sends an explicit request to the source for a new sequence number. This request is forwarded hop by hop to the source, with no regard to the feasibility condition. Upon receiving the request, the source increases its sequence number, and broadcasts an update, which is forwarded to the requesting node.

Note that after a change in network topology not all such requests will, in general, reach the source, as some will be sent over links which are now broken. However, if the network is still connected, then at least one among the nodes suffering from spurious starvation has an (unfeasible) route to the source; hence, in the absence of packet loss, at least one such request will reach the source. (Packet loss is compensated for by resending requests a small number of times.)

Since requests are forwarded with no regard to the feasibility condition, they may, in general, be caught in a forwarding loop; this is avoided by having nodes perform duplicate detection for the requests that they forward.

\subsection{Evaluation}
Babel has been reported to be a robust protocol and to have fast convergence properties\cite{abolhasan2010real}\cite{murrayexperimental}.

Babel offers high multi-hop bandwidth and fast route repair time. Babel has also been reported to outperform OLSR\cite{clausen2003optimized}, but it's still not suitable for highly mobile networks\cite{abolhasan2010real}.
Hence, the design of a high performance routing protocol for mobile ad hoc network (MANET) still remains an open research issue.

\section{How to solve the count to infinity problem}\label{sec:cti}
From the discussion of RIP protocol, we can see that routing loop and count to infinity problem is a demon that all distance vector routing algorithm must face. Next we will give some of our thoughts about the count to infinity problem and how to solve it. Later we'll examine some more sophisticated routing protocols and see how they cope with the problem.

\subsection{Thoughts}
A natural extension for Split-Horizon would be instead of storing the next hop information, we store the next $K$ hop in the routing table, and refuse to reveal the relevant routing information to them. However, this simple solution would not suffice. Suppose $K+2$ routers are directly connected to each other and one of them has a link to subnet $n$. If this link is broken and we would adopt the next-$K$-hop approach, a routing loop of $K+2$ would still form.

But, of course, if we set K to be $15$ in RIP, this approach would suffice, but anyway it's not a elegant solution and can't scale to larger networks. It is interesting to know that BGP protocol actually stores the information of all the Autonomous Systems the route has passed in the update message, and so in Wikipedia\cite{WikiDVRouting} someone claims that BGP is not a pure distance vector routing algorithm, and we will omit BGP in this survey.

\subsection{Our approach}
We can still work out a theoretical approach to prevent the routing loop problem in RIP protocol, and our approach is similar to some mechanisms in EIGRP, but its efficiency is still to be tested by practice. We call our approach RIP-Tree in this paper.

We consider the shortest path spanning tree to a particular subnet $n$. That is, if $B$ has $A$ as the next hop in its route to $n$, we add a edge from $B$ to $A$. Now it is easy to see the nodes (the routers that have a route to $n$) and these edges would form a tree. It is interesting to note that two routers, like $C$ and $D$ in the picture, although they have a direct link to each other, are pretty far away from each other in the tree.

\begin{figure}[htbp]
  \centering
  \includegraphics[width=150pt]{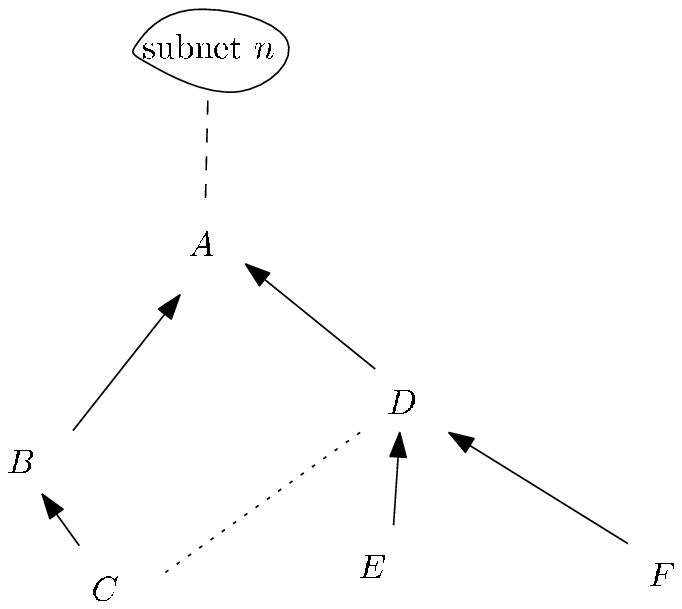}\\
  \caption{A typical shortest path spanning tree, where the arrow means parent-children relation and the dotted line means a direct link}\label{pic:RIP-TREE1}
\end{figure}

Now suppose a link or a router goes down. What does this mean to this shortest path spanning tree? It means a subtree  of nodes have to modify their route to n. In order to prevent routing loops in this process, we can first make all the nodes in the subtree to invalid their route to n, and then let the nodes figure out a new, valid route to n.

So how to make all the nodes in the subtree to invalid their route to $n$? We adopt a controlled flooding. Suppose $B$ detects that its link to $A$ is broken (perhaps by the loss of HELLO messages),and further $B$ has $A$ as the next hop to $n$. Now $B$ has to invalid all its descendants in the shortest path spanning tree to $n$.

First $B$ sends a unique INVALID\_ROUTE request to all its neighbors (except $A$, which is disconnected anyway, and with a unique sequence number to deal with duplicate messages), and all its neighbors would respond. In particular, those neighbors with $B$ as the next hop to $n$ would respond with a INVALID\_RELAY message, meaning that they will relay the message to all its descendants. $B$ will refuse to receive any updates about route to n until all its children has completed their tasks.

The process goes on in a recursive way until at the bottom of the spanning tree, a router with no children receives the INVALID\_ROUTE request, finds out it has completed the relay task, and informs its parent with a INVALID\_COMPLETE message. And this message goes all the way back to the root of the subtree. When the root receives INVALID\_COMPLETE message, it starts a controlled flooding again to spread the START\_ROUTING message. When a node receives a START\_ROUTING message, it starts to receive the update messages about $n$. To increase the stability of the algorithm, when a node sends out an INVALID\_ROUTE message, it starts a timer. When the timer expires, it also begin to receive updates. This feature is similar to hold-down in RIP protocol.

It is easy to understand why RIP-Tree is guaranteed to prevent routing loops. Note that when the root of the subtree receives a INVALID\_COMPLETE message, all the nodes of the subtree have invalid their route to n. After that the nodes will begin to receive updates about n, but none of the updates can originate from the subtree, so routing loop is prevented.
We should also consider the case when two or more subtrees perform this process simultaneously. Note that in this case all nodes of the subtree $B$ and $D$ will eventually find that they can't route to $A$. If node $C$, after the invalidation of subtree $B$, decides to route to $A$ through $D$. According to our design of RIP-Tree, $D$ has not started the invalidation process yet and will later inform $C$ that this route is also invalid.

Our algorithm also works best asymptotically. If the diameter of the network (the maximum distance between any two nodes) is $D$, it is easy to see that a routing update will take $D$ hops to reach the whole network in the worst case. By our method, we spend $2D$ hops to invalidate the subtree, so the algorithm is also $O(D)$. Our method suffers from the problem of single failure node (the root of the subtree), but we overcome this by setting a timer for each node.

\section{Conclusion}
In this section, we'll compare the five routing algorithm we've surveyed. RIP is by far the most popular and the simplest distance vector routing protocol, but it suffers from CTI problems. All other protocols have tried to solve this problem using different methods. Cisco's EIGRP protocol has been widely deployed on Cisco's routers. AODV is an on-demand ad-hoc mobile network routing protocol. RIP-MTI has the beautiful property of of downward compatibility with RIP protocol while avoiding CTI problem, and has little computation overhead. Babel integrates the advantages of EIGRP and DSDV, it uses sequence number to reduce starvation caused by SNC in EIGRP, and converges fast in ad hoc network. We give Table \ref{tab:comparison} comparing these five different routing algorithms.

We also put forward a new method, RIP-Tree, to eliminate the count to infinity problem. Our method is based on the spanning-tree property of the shortest paths in networks, and performs a more complex ``invalidation'' process (compared to RIP) when corrupted links are detected. RIP-Tree is an asymptotically fastest method, but its performance in practice is yet to be tested.

\begin{sidewaystable*}
\centering
\begin{tabular}{|p{2cm}|p{3cm}|p{3cm}|p{1cm}|p{3cm}|p{3cm}|p{3cm}|}
  \hline
Protocol Name & Scope of Application & Techniques & On-demand & Overhead & Convergence speed & Loop-free\\
  \hline
RIP & stable or slowly changing networks & reversed-poison, hold-down & NO & low(if there are no routing loops) & routing loops may greatly impede convergence & NO\\
  \hline
RIP-MTI & interior gateway routing with frequent topology changes & feasibility condition with loop detection, hold-down & NO & low & normal mode converges faster but may lead to routing loops, strict mode converges slower but is safer & strict mode is loop free, normal mode converges faster but is not loop free\\
  \hline
AODV & ad-hoc mobile networks & sequence number, controlled flooding of routing requests & YES & save overhead because of on-demand routing, but may cause more overhead in stable networks & fast & YES\\
  \hline
EIGRP & Interior gateway routing with frequent topology changes & DUAL, Feasibility Condition: SNC, Composite Metric & NO & low & fast & YES\\
  \hline
Babel & Wireless mesh networks and classical wired networks & Feasibility Condition: SNC, Sequence Number & NO & low & fast(presumably faster than EIGRP) & YES\\
  \hline
\end{tabular}
\caption{Comparison of RIP, RIP-MTI, AODV, EIGRP and Babel protocols}\label{tab:comparison}
\end{sidewaystable*} 
\newpage
\bibliography{library}

\begin{thebibliography}{10}

\bibitem{RIPGuide}
{The TCP/IP Guide - TCP/IP Routing Information Protocol (RIP, RIP-2 and
  RIPng)}.

\bibitem{abolhasan2010real}
M.~Abolhasan, B.~Hagelstein, and J.C.P. Wang.
\newblock {Real-world performance of current proactive multi-hop mesh
  protocols}.
\newblock In {\em Communications, 2009. APCC 2009. 15th Asia-Pacific Conference
  on}, pages 44--47. IEEE, 2010.

\bibitem{albrightson1994eigrp}
R.~Albrightson, JJ~Garcia-Luna-Aceves, and J.~Boyle.
\newblock {EIGRP-a fast routing protocol based on distance vectors}.
\newblock In {\em Proc. Networld/Interop}, volume~94, 1994.

\bibitem{bohdanowicz2009avoidance}
F~Bohdanowicz, H~Dickel, and C~Steigner.
\newblock {\em {Avoidance of Routing Loops}}.
\newblock Inst. f\"ur Informatik, 2009.

\bibitem{clausen2003optimized}
T.~Clausen, P.~Jacquet, C.~Adjih, A.~Laouiti, P.~Minet, P.~Muhlethaler,
  A.~Qayyum, and L.~Viennot.
\newblock {Optimized link state routing protocol (OLSR)}.
\newblock 2003.

\bibitem{garcia1993loop}
JJ~Garcia-Lunes-Aceves.
\newblock {Loop-free routing using diffusing computations}.
\newblock {\em IEEE/ACM Transactions on Networking (TON)}, 1(1):130--141, 1993.

\bibitem{Hu2003}
Y~Hu.
\newblock {SEAD: secure efficient distance vector routing for mobile wireless
  ad hoc networks}.
\newblock {\em Ad Hoc Networks}, 1(1):175--192, July 2003.

\bibitem{murrayexperimental}
D.~Murray, M.~Dixon, and T.~Koziniec.
\newblock {An Experimental Comparison of Routing Protocols in Multi Hop Ad Hoc
  Networks}.

\bibitem{Perkins2003}
C~Perkins and E~Belding-Royer.
\newblock {RFC3561: Ad hoc on-demand distance vector (AODV) routing}.
\newblock {\em Internet RFCs}, pages 1--38, 2003.

\bibitem{perkins1994highly}
C.E. Perkins and P.~Bhagwat.
\newblock {Highly dynamic destination-sequenced distance-vector routing (DSDV)
  for mobile computers}.
\newblock {\em ACM SIGCOMM Computer Communication Review}, 24(4):234--244,
  1994.

\bibitem{sklyarenko2006aodv}
G~Sklyarenko.
\newblock {AODV routing protocol}.
\newblock In {\em Seminar in Technische Informatik, Freie University, Berlin
  Germany}, 2006.

\bibitem{Steigner2008}
Ch. Steigner, H.~Dickel, and T.~Keupen.
\newblock {RIP-MTI: A New Way to Cope with Routing Loops}.
\newblock {\em Seventh International Conference on Networking (icn 2008)},
  pages 626--632, April 2008.

\bibitem{WikiDVRouting}
Wikipedia.
\newblock Distance-vector routing protocol --- {W}ikipedia{,} the free
  encyclopedia, 2010.
\newblock [Online; accessed 1-Dec-2010].

\bibitem{wiki:eigrp}
Wikipedia.
\newblock Enhanced interior gateway routing protocol --- {W}ikipedia{,} the
  free encyclopedia, 2010.
\newblock [Online; accessed 22-Nov-2010].

\end{thebibliography}
\end{document}